\documentclass[%
 reprint,
superscriptaddress,
 amsmath,amssymb,
 aps,
]{revtex4-2}
\usepackage{graphicx}
\graphicspath{{images/}}
\usepackage{dcolumn}
\usepackage{bm}
\usepackage[table,xcdraw]{xcolor}
\usepackage{booktabs}
\newcommand\hl[1]{%
  \bgroup
  \hskip0pt\color{red!80!black}%
  #1%
  \egroup
}
\usepackage{hyperref}
\usepackage{siunitx}

\DeclareSIUnit\Gbps{Gbps}

\begin{document}

\preprint{APS/123-RTD}

\title{Resonant tunnelling diode nano-optoelectronic spiking nodes for neuromorphic information processing}

\author{Mat\v{e}j Hejda}
\affiliation{Institute of Photonics, SUPA Dept of Physics, University of Strathclyde, Glasgow, United Kingdom}
\author{Juan Arturo Alanis}%
\affiliation{Institute of Photonics, SUPA Dept of Physics, University of Strathclyde, Glasgow, United Kingdom}
\author{Ignacio Ortega-Piwonka}
\affiliation{Dept de Física and IAC-3, Universitat de les Illes Balears, Palma de Mallorca, Spain}
\author{Jo\~{a}o Louren\c{c}o}
\affiliation{Centra-Ci\^{e}ncias and Departamento de F\'{i}sica, Faculdade de Ci\^{e}ncias, Universidade de Lisboa, Lisboa, Portugal}
\author{Jos\'{e} Figueiredo}
\affiliation{Centra-Ci\^{e}ncias and Departamento de F\'{i}sica, Faculdade de Ci\^{e}ncias, Universidade de Lisboa, Lisboa, Portugal}
\author{Julien Javaloyes}
\affiliation{Dept de Física and IAC-3, Universitat de les Illes Balears, Palma de Mallorca, Spain}
\author{Bruno Romeira}
\affiliation{Ultrafast Bio and Nanophotonics Group, International Iberian Nanotechnology Laboratory, Braga, Portugal}
\author{Antonio Hurtado}
\affiliation{Institute of Photonics, SUPA Dept of Physics, University of Strathclyde, Glasgow, United Kingdom}

\date{\today}

\begin{abstract}
In this work, we introduce an interconnected nano-optoelectronic spiking artificial neuron emitter-receiver system capable of operating at ultrafast rates ($\sim$ \SI{100}{\pico\second}/optical spike) and with low energy consumption ($<$ pJ/spike). The proposed system combines an excitable resonant tunnelling diode (RTD) element exhibiting negative differential conductance, coupled to a nanoscale light source (forming a master node) or a photodetector (forming a receiver node). We study numerically the spiking dynamical responses and information propagation functionality of an interconnected master-receiver RTD node system. Using the key functionality of pulse thresholding and integration, we utilize a single node to classify sequential pulse patterns and perform convolutional functionality for image feature (edge) recognition. We also demonstrate an optically-interconnected spiking neural network model for processing of spatiotemporal data at over \SI{10}{\Gbps} with high inference accuracy. Finally, we demonstrate an off-chip supervised learning approach utilizing spike-timing dependent plasticity for the RTD-enabled photonic spiking neural network. These results demonstrate the potential and viability of RTD spiking nodes for low footprint, low energy, high-speed optoelectronic realization of neuromorphic hardware.
\end{abstract}

\maketitle


\section{Introduction}
With the magnitude of data production increasing exponentially, machine learning (ML) approaches and the field of artificial intelligence (AI) have been undergoing a booming development, rapidly becoming ubiquitous in all domains of human endeavour. These methods have allowed machines to gain human-like information processing capabilities (e.g. learning, computer vision, natural language processing (NLP) or complex pattern recognition) and solve significant computational problems \cite{Callaway2020}. While AI algorithms achieve new breakthroughs, the hardware used to run those receives in turn less attention. Nowadays, large scale ML models are typically trained on cloud-based computing clusters, with some estimates placing the training energy consumption for a state-of-the-art NLP model on par with six years of total power energy consumption of a human brain \cite{Markovic2020}. Driven by the goal of reducing energy consumption as well as by the plateauing of empirical chip scaling laws, there has recently been significant growth of interest in non-conventional computing approaches. Neuromorphic (brain-like) engineering develops computer hardware architectures inspired by the brain and by the behaviour of biological neurons. Neuromorphic systems can be operated at various degrees of biological plausability, directly mapping conventional artificial neural network algorithms onto hardware or capitalising on the rich dynamical behaviour of biological neurons for information processing. While there already are powerful neuromorphic systems based on electronics \cite{DeBole2019,Davies2018}, the reliance on CMOS technology imposes limits in terms of interconnectivity and component density, with dozens of transistors required per neuron and additional external memories needed for synaptic weights. This results in several $\mu m$ large neurons. Since dedicated wiring for every synaptic link is not practical, neuromorphic electronic systems usually employ a shared digital communication bus with time-division multiplexing \cite{Merolla2014}, gaining interconnectivity at the expense of bandwidth, or use schemes such as address-event representation (AER) \cite{Pfeiffer2018}. As an alternative, hardware technologies relying on physics for neuromorphic computation are nowadays gaining increasing research interest. These include hybrid CMOS/memristive systems (see \cite{Markovic2020} for an overview), spintronics \cite{Grollier2020} and photonic systems \cite{Shastri2021, Sunny2021}. 
\begin{table*}[ht!]
\caption{\label{tab:comparison}Comparison of photonic and optoelectronic technologies capable of spike (pulse) based signalling.}
\begin{tabular}{@{}lll@{}}
\toprule
\rowcolor[HTML]{ededed} 
Photonic platform                               & Energy/event                          & Spike   event timescales    \\ \midrule

superconducting Josephson junctions (cryogenic) \cite{Shainline2018}     & $>2\cdot10^{-14}$ J & $>\SI{1}{\nano\second}$ (nTron switching)  \\

phase change material cells \cite{Chakraborty2018}                              &  $\sim10^{-12}$ J           & $\sim\SI{500}{\pico\second}-\SI{1.5}{\nano\second} $ (read/write)           \\

micropillars \cite{Selmi2016}                     & $\sim5\cdot10^{-14}$ J (excl. pump)              & $\sim\SI{200}{\pico\second}$                 \\

graphene-SA laser \cite{Ma2018}                      & $\sim10^{-8}$ J                                     & $\sim\SI{20}{\micro\second}$                       \\
RTD optoelectronic node [this work]                                      &$\sim10^{-13}$ J                                    & $\sim\SI{100}{\pico\second}$         \\ \bottomrule
\end{tabular}

\end{table*}
Neuromorphic photonics is a nascent field, recently gaining significant traction due to increasing importance of AI algorithms and rapid advances in the field of photonic integrated circuits (PICs). Optoelectronic systems in particular are considered as highly suitable for future cognitive computing hardware, as they benefit from operation with both electrons and photons, each excelling at different key functionalities \cite{Shainline2018a}. Thanks to their capability to address bandwidth and interconnect energy limits in a scalable fashion, optoelectronic systems might prove as the optimal solution to overcome these limitations \cite{Miller2017}. There are many different approaches to realization of artificial neural networks in optics (see for example review \cite{DeMarinis2019}). Using delayed feedback, recurrent neural networks can be realized in a photonic reservoir computer, yielding networks with large number of virtual nodes while requiring only very low hardware complexity \cite{Bueno2018}. Closer to the usual digital implementation of artificial neural networks are platforms that enable accelerated matrix/tensor-based computation \cite{Shen2017, Feldmann2021}. Some photonic systems, such as diffractive surfaces \cite{Lin2018, Luo2019}, could allow for passive computation by interaction between light and matter. One of the key principles when designing biologically-plausible neuromorphic hardware is excitability and event-based signalling. Biological neurons communicate with electronic signals using a sparse encoding scheme known as spiking. Photonic neuromorphic systems that utilize spike signalling include phase-change material based integrated networks of micro-ring resonators \cite{Chakraborty2018,Feldmann2019}, photonic crystals \cite{Laporte2018}, superconducting Josephson junctions \cite{Shainline2018}, micropillar lasers \cite{Pammi2020}, excitable semiconductor lasers, including a graphene laser with saturable absorber \cite{Ma2017}, quantum-dot laser \cite{Mesaritakis2016, Robertson2018, Sarantoglou2020}, micro-ring resonators \cite{Xiang2020_MRR}, vertical cavity surface emitting lasers (VCSELs) \cite{Robertson2020a, Hejda2020,Hejda2021} and multi-section VCSELs with saturable absorber \cite{Xiang2020, Zhang2020}. Table \ref{tab:comparison} provides comparison of some of these approaches. This wide array of investigated technologies demonstrates the power and high potential of photonics for unconventional brain-inspired computing. Despite the impressive progress, the development of a single, miniaturized light-emitting nanoscale source and detector for spike-based operation (which is key for neuromorphic nano-optoelectronic computing) remains an ongoing, significant challenge.  
\begin{figure*}[t]
\centering
\includegraphics[width=0.99\textwidth]{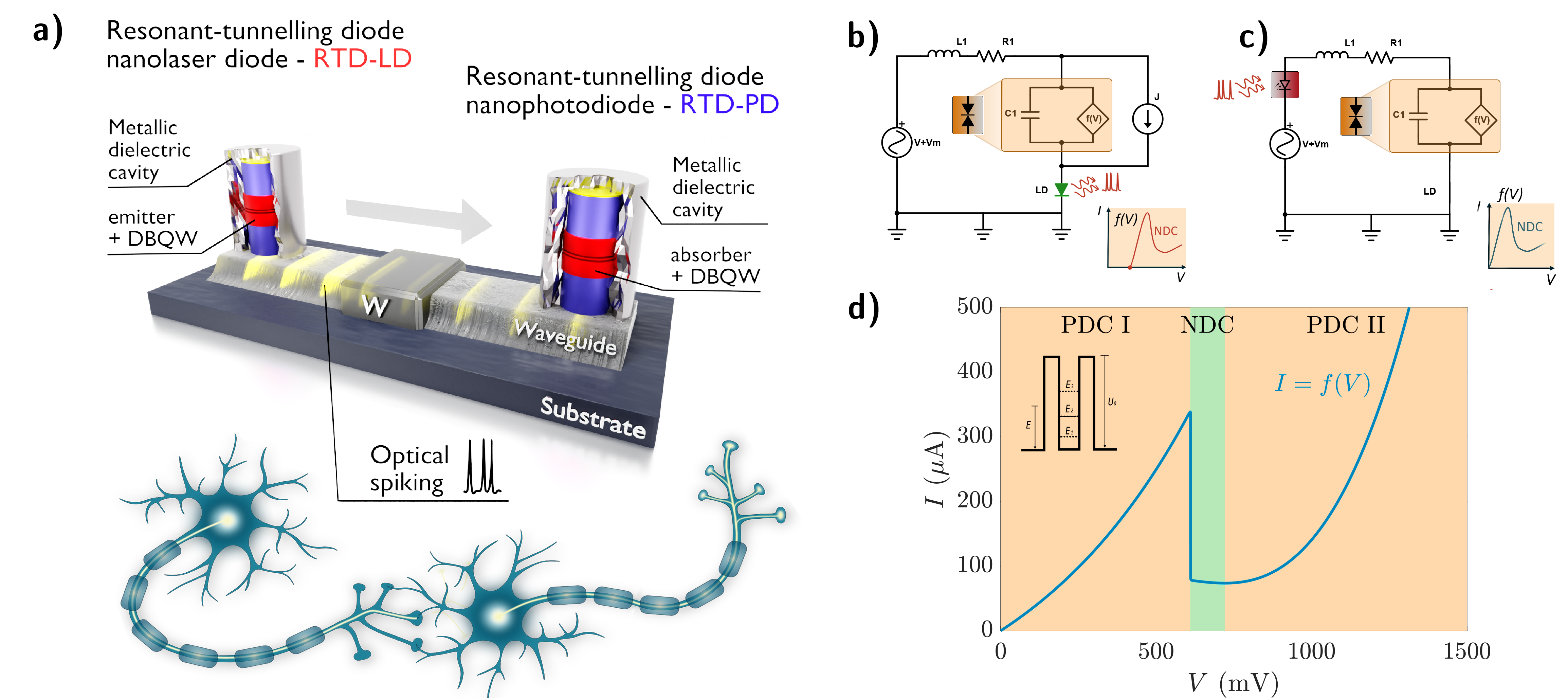}
	\caption{\label{fig:pillars}\emph{(a)} Illustration and circuit diagrams of the proposed solution for neuromorphic system based on two types of RTD-powered optoelectronic nodes: RTD-LD (master) and RTD-PD (receiver) nodes. The RTD-LD and RTD-PD metal-dielectric encapsulated micropillars are coupled using a waveguide with adjustable attenuation factor $W$. When subject to external bias, RTD-LD nodes can respond to incoming perturbations with short optical pulses (spikes), which can be processed in the downstream RTD-PD node. This functionality mimics the use of action potential in biological neurons. \emph{(b)} Lumped circuit scheme for the RTD-LD node. \emph{(c)} Lumped circuit diagram from the RTD-PD node. \emph{(d)} The RTD I-V characteristic used in this study, with curve parameters obtained by fitting experimental data. Regions of positive differential conductance (PDC) and negative differential conductance (NDC) are highlighted in colour. Inset shows a simplified DBQW scheme with the discrete energy levels. Typical thickness of the DBQW region is around \SI{10}{\nano\metre}}
\end{figure*}

\section{NEUROMORPHIC RTD-POWERED OPTOELECTRONIC NODES}
In this work, we introduce an optoelectronic neuromorphic system utilizing a resonant tunnelling diode (RTD) element based on a double barrier quantum well (DBQW) epi-layer structure. The DBQW consists of a narrow bandgap semiconductor layer embedded between two thin layers with wider bandgap (Fig. \ref{fig:pillars}d, inset), with typical barrier thicknesses ranging from 4 to \SI{8}{\nano\metre}, and \SI{1}{\nano\metre} to \SI{3}{\nano\metre}, respectively. Under applied voltage, the structure works as a filter for the carrier’s energy, leading to high carriers’ transmission when energy of the electrons (Fermi sea) resonates with the confinement energy levels of the DBQW. The voltage-controlled probability for incident electrons to cross the barrier is locally maximized, which results in the typical N-shaped function voltage-current relation $f(V)$ with one or more regions of negative differential conductance (NDC) in between two or more regions of positive differential conductance (PDC) \citep{WAW-IEEE-18} as shown in Fig. \ref{fig:pillars}d. The presence of the nonlinearity and gain in the NDC region, persisting from DC up to THz frequencies \cite{Izumi2017}, makes RTDs particularly suitable for high frequency oscillators \cite{Feiginov2019}. The nonlinearity is key for operation of the proposed neuromorphic RTD node as a fast, excitable spiking nonlinear source \cite{Ortega-Piwonka2021} with intrinsic electrical gain. Previous works have investigated triggering of stochastic excitable responses in hybrid integrated optoelectronic RTD circuits \cite{Romeira2013, Romeira2014} and operation of RTDs with delayed feedback \cite{Romeira2017}, addressing only operation of a single (solitary) device. In this work, we investigate for the first time interconnected systems consisting of multiple independent RTD-based monolithic integrated optoelectronic nodes. We employ the nodes as stateless excitable devices and take advantage of the spike-based signalling to implement information processing tasks and multi-device networks, with prospects for very low footprint, low energy and high-speed operation due to sub-$\lambda$ elements. We utilize two types of nodes: an electronic-optical (E/O) RTD-LD system, realized with a RTD element coupled to a nanoscale laser diode (LD), and a optical-electronic (O/E) RTD-PD system, realized with a photodiode (PD) coupled to a RTD element. In both node types, spiking threshold can be adjusted via bias voltage tuning. An illustration of two nodes with an unidirectional optical weighted link, representing two feedfoward linked neurons, is depicted in Fig. \ref{fig:pillars}a. 
\subsection{Optoelectronic RTD-system architecture}
In both the RTD-LD and RTD-PD nodes, the two functional blocks are integrated in a monolithic, metal dielectric cavity micro-pillar with the DBQW regions on GaAs/AlGaAs materials \cite{Romeira2020} for operation at the wavelength of \SI{850}{\nano\metre} and InP materials \cite{Romeira2013a} for operation at \SI{1550}{\nano\metre}. For simplicity, in this work we focus on one of the two material platforms and investigate InP-based RTD systems throughout our analyses. The micro-pillar is coated by a dielectric cap (typically made of SiO\textsubscript{2}) and metallic layer (typically Au or Ag), similarly to previously reported waveguide-coupled nanoLEDs \cite{Dolores-Calzadilla2017}. A significant advantage of the semiconductor RTD epilayer design is that it can be used to realize all the required functional blocks in the proposed neuromorphic optoelectronic nodes, including ultra-sensitive photodetectors \cite{Pfenning2015,Pfenning2016}, high bandwidth nonlinear behaviour (including spiking responses) in the electric domain, and light emission, including both coherent (laser) and non-coherent (light emitting diodes, LEDs) operation. This brings the possibility of all-in-one monolithic integration of the required functional blocks into singular sub-micron scale devices. Specific epi-layer designs based upon different materials platforms targeting operation at forementioned wavelength ranges, i.e. \SI{1550}{\nano\metre} (InP) and \SI{850}{\nano\metre} (GaAs), are currently being investigated towards the fabrication of the systems proposed in this work. For non-coherent signalling between nodes, the RTD-LD can also be realized using an RTD-LED sub-$\lambda$ element at high (multi-gigahertz) speeds with very low power consumption (\SI{<1}{\pico\joule} per emitted spike) \cite{Romeira2020}. It was observed that the light emission efficiency of the pillar design increases with smaller sizes, with sub-lambda pillars yielding very high light-extraction efficiency \cite{Romeira2020_eff}. RTD-powered nanolasers and light-sources may also benefit from their small size in terms of improved operation speed and reduced lasing threshold \citep{Romeira2020_limits}. In a review \cite{Markovic2020}, it was stated that a minimum lateral size of hardware neurons is to be expected around \SI{100}{\micro\metre}. RTD components, embedded as singular or monolothic sub-micron structures, have the potential to be significantly smaller, overcoming one of the key expected disadvantages (large footprint) in such systems. Unlike the superconducting and fluxonic \cite{Shainline2020} solutions, the RTD-based optoelectronic node can be operated at room temperatures.

The synaptic links in this work are required for optical signal propagation between nodes and signal weighting (controllable optical signal attenuation). Recent advances in integrated, tuneable waveguide meshes offer chip-scale solutions for linear matrix transformations \cite{Perez2018}, which typically underpin the weighting functionality in neural networks. The micropillars can be directly coupled to waveguides by heterogenous integration \cite{Jevtics2020} or coupled together by means of two-photon polymerization \cite{Maibohm2020} guiding structures. Signal attenuation in photonic waveguides can be realized for example by means of balanced Mach-Zehnder interferometers, directional couplers \cite{Perez2018} or nano-scale phase change material (PCM) cells \cite{Michel2020}. PCM-based synaptic cells also exhibit suitability for all-optical spike-timing based plasticity \cite{Cheng2017} and all-optical synaptic signal weighting can also be realized using VCSOAs \cite{Alanis2021}. Synaptic interconnections can also be realized using photorefractive III-V photonic structures on silicon \cite{Stark2020}.
\section{RTD-LD$\rightarrow$RTD-PD: Theory and Dynamics}\label{sec:Theory_and_Dynamics}

\subsection{Single optoelectronic node\label{subsec:node}}
We consider the monolithic nodes as optoelectronic circuits based on an RTD element connected to electrical and/or optical modulation (Fig. \ref{fig:pillars}(b,c)). The circuit dynamics are described by Kirchhoff laws, together with a nanolaser diode model \citep{BF-IEEE-18,YB-JAP-89,ML-APL-18,BY-IEEE-91,RC-PRA-94}: 
\begingroup
\allowdisplaybreaks
\begin{align}
C\frac{dV}{dt} & =I-f(V)-\kappa S_{m}(t)\label{eq:RTD_V}\\
L\frac{dI}{dt} & =V_{m}(t)-V-RI\label{eq:RTD_I}\\
\frac{dS}{dt} & =\left(\gamma_{m}(N-N_{0})-\frac{1}{\tau_{p}}\right)S+\gamma_{m}N+\sqrt{\gamma_{m}NS}\xi(t)\label{eq:LD_S}\\
\frac{dN}{dt} & =\frac{J+\eta I}{q_{e}}-(\gamma_{l}+\gamma_{m}+\gamma_{nr})N-\gamma_{m}(N-N_{0})|E|^{2}\label{eq:LD_N}
\end{align}
\endgroup
\noindent Here, $V$ is the voltage along the RTD, $I(t)$ is the circuit's total current, $S(t)$ is the photon number and $N(t)$ is the carrier number. $R$ is the circuit equivalent resistance and $L$ is the intrinsic inductance of the circuit while $C$ is the parasitic capacitance of the RTD. $V_{m}(t)$ is the modulation voltage function. We consider two node models: a) an O-E RTD integrated with a photodetector (PD), governed by Eqs. \ref{eq:RTD_V}-\ref{eq:RTD_I}, which can be driven by external optical pulses (represented as $S_{m}(t)$) where $\kappa$ is the photodetector conversion factor translating input optical intensity into a photocurrent \cite{Romeira2017} signal; b) an E-O RTD-LD node, governed by all the shown equations (Eq. \ref{eq:RTD_V}-\ref{eq:LD_N}) with omission of the PD term. We assume low input optical power level (with small power variations), allowing for use of linearized sensitivity-power relation in the PD term \cite{Pfenning2016} and static $f(V)$. Due to reduced cavity size, the spontaneous and stimulated emission rates are modified as a result of Purcell enhancement of both the radiative processes \cite{BF-IEEE-18}. For simplicity of analysis, the rate equation model includes only homogeneous broadening effects. $N_{0}$ is the transparency carrier number, $\text{\ensuremath{\tau_{p}}}$ is the photon lifetime, $\gamma_{m},\gamma_{l},\gamma_{nr}$ are respectively the spontaneous emission rate into the lasing mode (where $\gamma_{m}*S$ is the stimulated emission rate), radiative decay into the leaky modes and non-radiative spontaneous emission coefficients. $q_{e}$ is the electron charge and $J$ is an input bias current injected into the LD in addition to the RTD current $I(t)$. The stochastic nature of the system is given in Eq. \ref{eq:LD_S} by the term $\gamma_{m}N$ and the multiplicative noise $\sqrt{\gamma_{m}NS}\xi(t)$, where $\xi(t)$ is a time-uncorrelated white noise function. The parameters used in this work are available from the Supplementary information. The function $f(V)$ accounts for the nonlinear relation between the voltage applied across the RTD and the current passing through it. We use an analytical expression for $f(V)$ derived in \citep{SDC-EDL-96} and detailed in the Supplementary Information. The device operates at room temperature (300°K). Fig. \ref{fig:pillars}b shows the experimentally fitted I-V characteristic (parameters available from Supplementary information) with a relatively narrow region of negative differential conductance embedded in between two regions of positive differential conductance, labelled as NDC, PDC I and PDC II, respectively. The curve peak is located at $V$ = \SI{609.6}{\milli\volt}, with a local maximal current of \SI{338.6}{\micro\ampere}. At the right of the peak, the current abruptly drops from \SI{340}{\micro\ampere} to \SI{80}{\micro\ampere} in a span of less than \SI{1}{\milli\volt}. Further rightwards, $f\left(V\right)$ continues to decrease although with a much more moderate rate, until it reaches a valley at $V$ = \SI{720.7}{\milli\volt} and a local minimal current of \SI{73.6}{\micro\ampere}. Beyond this point, $f\left(V\right)$ increases again following a diode-like behaviour.

\subsection{Dynamical behaviour\label{subsec:Optoelectronic_integration}}
\begin{figure}[t]
\centering{}
\includegraphics[width=0.91\columnwidth]{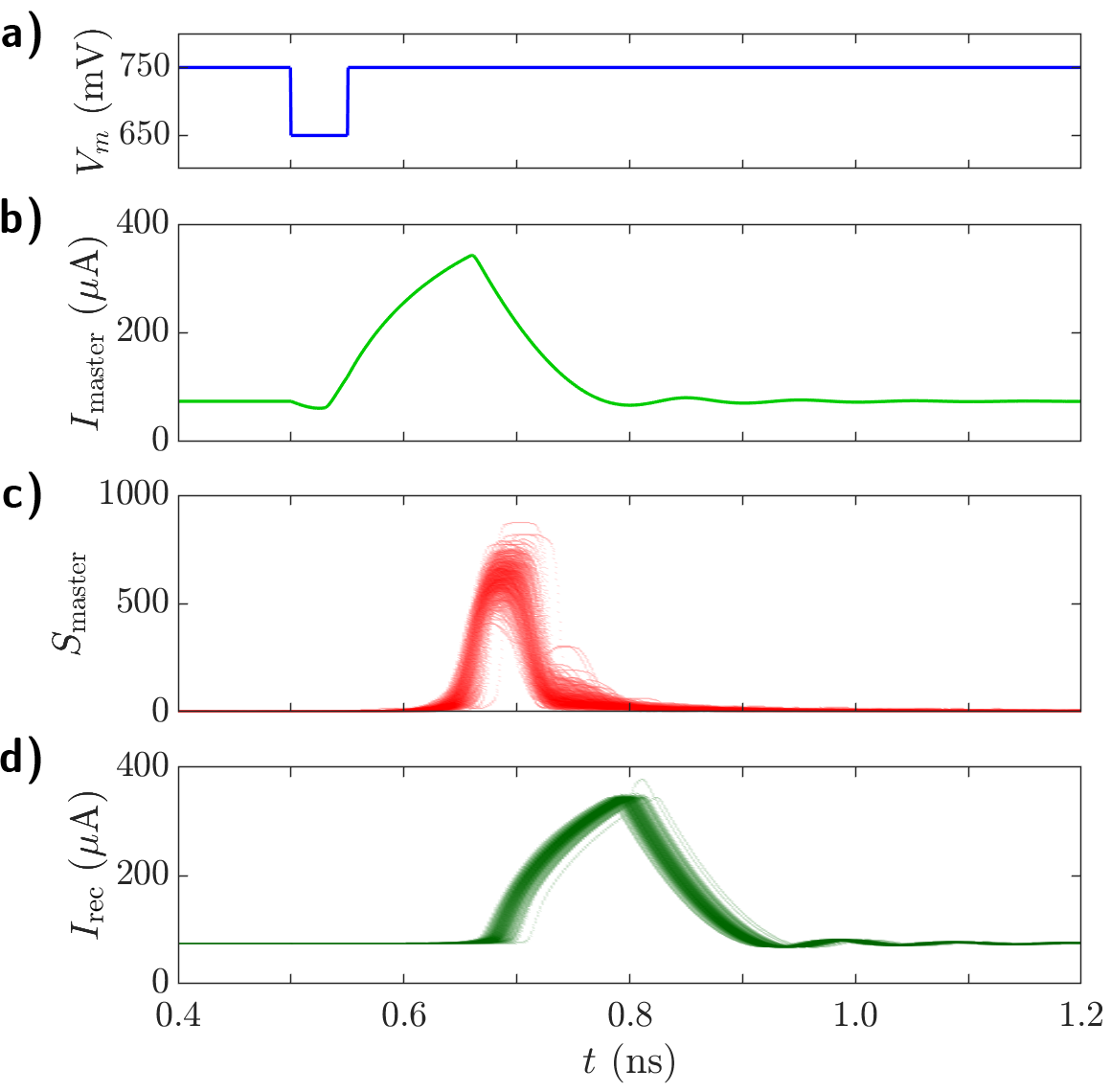}
\caption{Steps in 500 responses of the master-receiver optoelectronic system to the same input square pulse. The RTD elements and LDs are biased at $V_{0}$ = \SI{750}{\milli\volt} and $J$ = \SI{210}{\micro\ampere}, respectively. \emph{a}) Square voltage pulse injected into the master RTD element. \emph{b}) Master RTD electronic pulse response. \emph{c}) Master LD optical pulse response. \emph{d}) Receiver RTD element electronic pulse response.}\label{fig:Master_Slave_outputs}
\end{figure}

When the system (Eqs. \ref{eq:RTD_V},\ref{eq:RTD_I}) is biased in the proximity of the peak or valley of its I-V curve and injected with positive or negative voltage pulses respectively, it behaves as an excitable system able to respond with electronic spikes. Using this functionality, numerical simulations of Eqs. (\ref{eq:RTD_V},\ref{eq:RTD_I},\ref{eq:LD_S},\ref{eq:LD_N}) are run, where a train of square negative voltage pulses $V_m$ is used to trigger a spike in the RTD-LD optoelectronic master node (Fig. \ref{fig:Master_Slave_outputs}\emph{a}). The period of the train is \SI{2}{\nano\second} and each pulse is \SI{50}{\pico\second} long and \SI{100}{\milli\volt} deep. No optical modulation is used (i.e., $S_{m}(t)=0$). The RTD is biased close to the valley point at \SI{750}{\milli\volt}. 50 simulations are run over 10 periods (thus, a total of 500 pulses are injected). The RTD responds with upwards current pulses (Fig. \ref{fig:Master_Slave_outputs}\emph{b}), each about \SI{275}{\pico\second} long and reaching a peak of \SI{342}{\micro\ampere}. The response delay is roughly \SI{25}{\pico\second} and the rest value of the signal is \SI{74}{\micro\ampere}. Such a pulse elicits a weak response when injected into the LD because its peak value it slightly surpasses the LD threshold current for a very brief time. This is why the additional input bias current $J$ is necessary. When the LD is biased at $J$ = \SI{210}{\micro\ampere}, the total current injected $J+I_{master}(t)$ has a rest value of \SI{284}{\micro\ampere} and a peak value of \SI{552}{\micro\ampere}, well below and well above the threshold current, respectively. In consequence, the LD remains inactive most of the time and emits a pulse in response to each current pulse (Fig. \ref{fig:Master_Slave_outputs}\emph{c}). The optical pulse is shorter, with a duration of 40-\SI{60}{\pico\second} (with temporal fluctuations due to the white noise term in Eq. \ref{eq:LD_S}) because the LD takes a relatively long time to respond to an above-threshold current, while it quickly stops emitting as the current descends under the threshold. This results in the optical pulse being shortened and the response latency increased up to about \SI{75}{\pico\second}. This phenomenon is typical in systems that exhibit transcritical bifurcations and is known as \emph{critical slowing down} \citep{TLP-AJP-04,MAB-PRL-20,SSB-OptComm-87,ME-PRL-84}. 
The estimate for RTD-LD power consumption is based on the idle state current (\SI{284}{\micro\ampere}), multiplied by the idle voltage bias (\SI{750}{\milli\volt} for valley point), resulting in \SI{213}{\micro\watt}. This is inclusive of the additional $J$ term that sets the sub-threshold operation current of the laser. The spiking itself, due to its very short temporal timescales, will require small amount of additional power. Assuming for the spiking event a peak current of \SI{552}{\micro\ampere} and same voltage value of \SI{750}{\milli\volt} gives a power of \SI{414}{\micro\watt} during an approximate time of \SI{100}{\pico\second} (based on pulse shape from Fig. \ref{fig:Master_Slave_outputs}b) with maximum spiking repetition rate interval of approx. \SI{420}{\pico\second}. Hence, the upper bound on power consumption in the system can be taken as temporally weighted average of the \textit{spike} (\SI{100}{\pico\second}) and \textit{idle} (\SI{320}{\pico\second}) states, resulting in \SI{261}{\micro\watt}. Higher firing sparsity (lower spiking rate) below the maximum achievable inter-spike timing interval will reduce the total power consumption. With upper bound on spike firing repetition rate of \SI{420}{\pico\second}, the total energy consumption per spike can reach values as low as \SI{110}{\femto\joule}. We also note that peak and valley voltages in RTDs can be much smaller than 0.5 V, and that RTDs and nanolasers can be designed for operation at lower currents (\SI{10}{\micro\ampere} – \SI{100}{\micro\ampere}) \cite{Romeira2018} to further reduce power consumption. In summary, the optoelectronic RTD-LD node has been demonstrated as an excitable system able to generate short optical pulses with low power consumption.
To facilitate networking, the optical pulse leaving the optoelectronic node can be used to drive a second node in a master-receiver integrated circuit. With the receiver RTD-PD circuit biased close to the valley of its I-V characteristic ($V_{m}(t) = V_{0} = $ \SI{750}{\milli\volt}), the perturbation $\kappa S_{\text{master}}(t)$ is able to elicit an excitable response from the receiver RTD in the form of an excitatory current pulse similar to that produced by the master RTD (Fig. \ref{fig:Master_Slave_outputs}\emph{d}), albeit with a fluctuating character. Therefore, we show the master-receiver integrated circuit is able to propagate (cascade) information by means of optical pulses. The low required values of the $\kappa$ conversion factor (see Supplementary Information) used in the model demonstrate that cascaded responses require only a small portion of the optical output energy produced by upstream nodes, further increasing the prospects of larger fan-in/fan-outs in networks.  
\section{Information processing with RTD-based optoelectronic nodes}

\subsection{Single node 8-bit pattern recognition task}

\begin{figure*}[ht]
\centering
\includegraphics[width=0.78\textwidth]{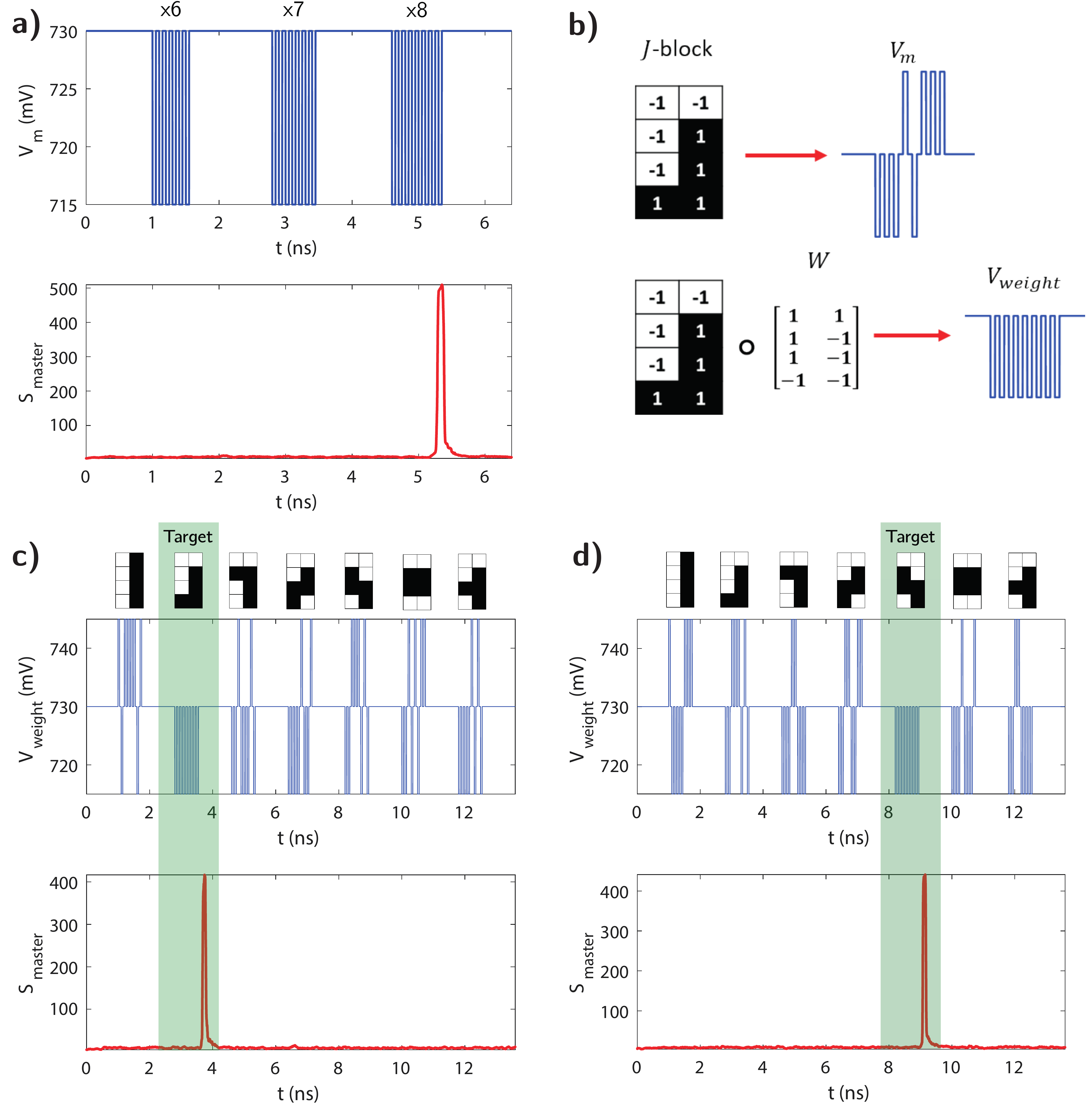}
	\caption{\emph{a)} RTD-LD response to an AC modulation signal: containing three sets of negative square signals with 6, 7 and 8 negative pulses respectively (top) and corresponding LD output trace (bottom). \emph{b)}  Example of a Tetris $J$-block represented by a 4$\times$2 grid and corresponding serialized signal $V_{m}$ (top). The $J$-block is weighted offline by element-wise multiplication with a matrix $W$ converting $V_{m}$ to $V_{weight}$ (bottom). \emph{c)}-\emph{d)} Simulation of pattern recognition tasks, where $W$ was chosen to target the $J$ \emph{c)} and $S$-block \emph{d)} respectively. The corresponding driving signal $V_{weight}$ is shown (top) accompanied by the LD output trace (bottom). The LD outputs have been smoothed by taking a moving average $t_{MA} =$\SI{0.1}{\nano\second} to approximate the effect of the response time of the photodetector and to ease the visualization.}\label{fig:integrate_single}
\end{figure*}

Neurons have the ability to integrate a series of input stimuli and elicit a single spike firing response. This happens due to the cumulative effect of separate input perturbations which, when combined, can exceed the neuron firing threshold intensity. A similar integrate and fire (I\&F) behaviour can be replicated with RTD devices. 

To demonstrate this, we modelled the dynamical response of a single RTD-LD node when driven by an AC signal $V_m$ consisting of short negative sub-threshold square pulses. In this case, the RTD was biased at a voltage $V_{DC}$ = \SI{730}{\milli\volt} ($I_{DC}$ = \SI{73}{\micro\ampere}), which positions the device's operation point in the valley slightly to the right of the NDC region. The LD was biased at $J$ = \SI{210}{\micro\ampere}, thus the total current injected ($J+I_{DC}$) has a rest value of \SI{283}{\micro\ampere} (below the firing threshold current). For simplicity, we do not include a receiver RTD-PD circuit, but it is assumed that a perturbation $S_{master}$ can be propagated to a receiver node in the form of an excitatory current pulse. To show the circuit's I\&F functionality, the RTD element was driven by an AC signal consisting of \SI{50}{\pico\second} pulses of amplitude $V_{ac}$ = \SI{-15}{\milli\volt}, separated by \SI{50}{\pico\second}. Thus, the resulting modulation signal is $V_{m}$=$V_{DC}+V_{ac}$. Figure \ref{fig:integrate_single}a shows the input signal $V_m$ (top), which consists of three pulse trains with $\times$\,6, $\times$\,7 and $\times$\,8 pulses respectively, and the resulting RTD-LD output trace $S_{master}$ (bottom) as a function of time. For modulation signals containing $<$\,8 pulses the output remains unperturbed. However, as the number of pulses is increased to 8, their combined intensities trigger a firing event in the RTD element, eliciting in turn a spike in the LD output. This I\&F behaviour can be exploited to perform 8-bit pattern recognition tasks by a single RTD-LD node at very high speed rates, as it is demonstrated in Fig. \ref{fig:integrate_single}b-d.

In this example, seven different 8-bit patterns, representing Tetris-like blocks, are mapped into a 4 by 2 grid with individual values of 1 and -1 (representing black and white colour pixels respectively). The corresponding pattern is described by $V_{m}$ as a serialized 8-bit signal (top of Figure \ref{fig:integrate_single}b). Subsequently, each pattern is multiplied offline by an array of weights $W$ associated to a target Tetris piece. In the example shown in Figure \ref{fig:integrate_single}b, the element-wise multiplication between the $J$-block pattern and $W$ = $[-1, -1, -1, 1, -1, 1, 1, 1]$ converts the input to a serialized all-negative 8-bit signal $V_{weight}$. For the simulation, $V_{weight}$ included 7 patterns temporally separated by \SI{1}{\nano\second}. Each bit had an activation time of \SI{50}{\pico\second}  with an amplitude  $V_{ac} = \pm$\SI{15}{\milli\volt}, separated by \SI{50}{\pico\second}. Two examples of a weighted modulation signal, used to recognise a $J$-shaped and $S$-shaped target piece respectively, along with their corresponding LD ($S_{master}$) output traces, are shown in Figure \ref{fig:integrate_single}c-d. As highlighted by the shaded green boxes, the RTD-powered node is able to successfully integrate 8 bits and fire as indicated by a sharp pulse in the LD output; thus being able to recognise the desired target piece in each case.
\subsection{Image edge detection task using sub-threshold pulse integration}

\begin{figure*}[ht]
\centering
\includegraphics[width=0.89\textwidth]{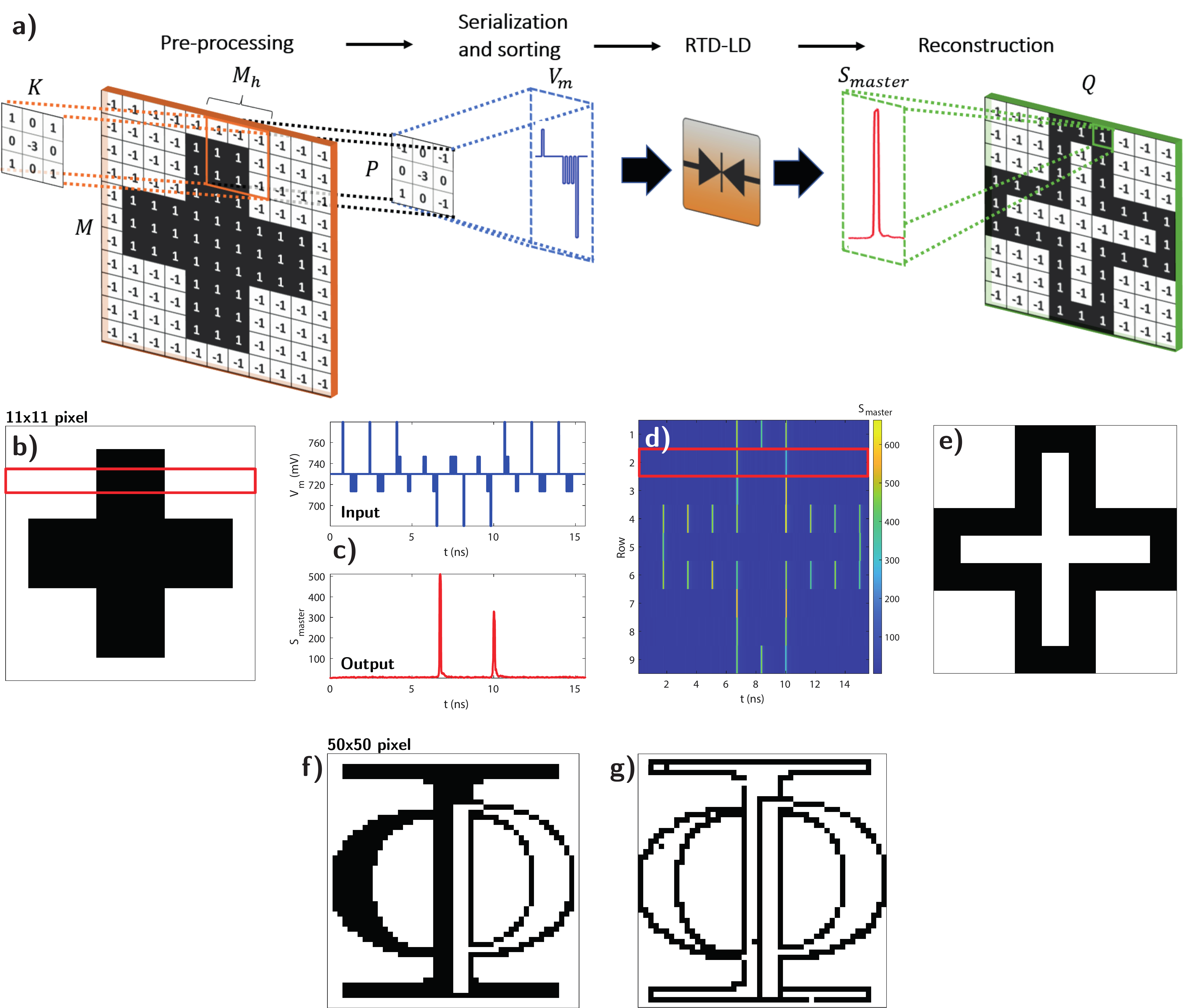}
	\caption{\emph{a)} Steps followed to perform an edge detection task by a RTD-LD device. The process consists of four main phases: offline multiplication of a binary image $M$ and kernel $K$, serialization and sorting of the 9-bit pattern to generate a modulation signal $V_{m}$,  simulation of RTD-LD  response to $V_{m}$, and reconstruction of the LD output to a binary image $Q$. \emph{b)} 11$\times$11 pixels binary image used for edge detection task, where pixels are assigned values of 1 (black) or -1 (white). \emph{c)} Example of modulation signal used as input to drive the RTD (top) and corresponding LD output trace (bottom). \emph{d)} Colour plot showing the complete LD output series used for edge detection of $M$ (The red box corresponds to the $S_{master}$ output plot shown in \emph{c)}. \emph{e)} Reconstruction of the LD output trace into a binary image $Q$. \emph{f)} 50$\times$50 binary image of Strathclyde's Institute of Photonics (IoP) logo. \emph{g)} Reconstructed image after single RTD-LD edge detection task.}\label{fig:edge_detect}
\end{figure*}

We further demonstrate the possibility of using a single RTD-LD node to perform image edge-detection operation. For this task we utilized a binary image $M$ of size $n \times n$ (Figure \ref{fig:edge_detect}a), where black and white pixels are assigned values of 1 or -1 respectively. On the pre-processing phase, an element-wise product between a 3 $\times$ 3 matrix kernel $K$ and sections of the binary image $M_h$ is performed offline such that:

\begin{multline}
    P = K \circ M_h  \\ = 
    \begin{bmatrix}
    1 & 0 & 1\\
    0 & -3 & 0\\
    1 & 0 & 1
    \end{bmatrix} \circ 
    \begin{bmatrix}
    M_{i,j} & M_{i,j+1} & M_{i,j+2}\\
    M_{i+1,j} & M_{i+1,j+1} & M_{i+1,j+2}\\
    M_{i+2,j} & M_{i+2,j+1} & M_{i+2,j+2}
    \end{bmatrix}
    \label{k_times_m}
\end{multline}

\noindent where $i$ and $j$ are the indices of the individual pixels in $M_h$. The resulting matrix $P$ is serialized as a 9-bit pattern, where each bit is assigned a \SI{50}{\pico\second} activation pulse and a \SI{50}{\pico\second} separation for a total of \SI{100}{\pico\second} per bit. Each pulse is assigned an amplitude $V_{ac} = \pm$\SI{16}{\milli\volt} $\ast P_{k,l}$, where $k$ and $l$ are the indices of individual matrix elements in $P$. The serialized bits are sorted such that their amplitude is rearranged in descending order. This ensures all negative pulses are integrated consecutively to elicit a firing response. The resulting 9-bit modulation signal ($V_{m}$) is used as the electrical input for the RTD-LD node. The process described above is repeated for each row of $M$ taking steps of 1 pixel. Finally, the output of the RTD-LD device is used to reconstruct a binary image $Q$, where pixels are assigned a value of 1 if the laser output trace exhibits a spike and -1 otherwise. 

An example of an 11$\times$11 binary image, used to demonstrate edge detection operation, is shown in Figure \ref{fig:edge_detect}b. Each row of $M$ is described by a modulation signal $V_{m}$, like shown in the top of Figure \ref{fig:edge_detect}c, consisting of 9 patterns with a duration of 100 $\times $ \SI{9}{\pico\second} each and temporally separated by \SI{750}{\pico\second} to account for the time required for the LD output to return to zero. For the simulation the RTD was biased at the valley $V_{DC}$ = \SI{730}{\milli\volt}. The corresponding $S_{master}$ time trace, displayed in the bottom of Figure \ref{fig:edge_detect}c, shows two spikes of the LD output (pixels 4 and 6) as a result of the I\&F response of the RTD (red box in Figure \ref{fig:edge_detect}b,d). Figure \ref{fig:edge_detect}d shows a colour plot of the LD output traces for each row of $M$, where the high values of $S_{master}$ correspond to a detected edge. A binary image $Q$, reconstructed from the RTD-LD output, is shown in Figure \ref{fig:edge_detect}e. It can be observed that, following an offline element-wise multiplication operation with a single 3 by 3 kernel, the RTD-LD node is able to consistently detect all edges of $M$, regardless of their orientation. We further show the capability of an RTD-LD to consistently detect all edge features, by using a 50$\times$50 pixels binary image of the logo of the Institute of Photonics (IoP) at the University of Strathclyde (Fig. \ref{fig:edge_detect}f). The reconstructed image in Figure \ref{fig:edge_detect}g shows the RTD-LD node is able to detect all edges with a 99.7\% accuracy. This results are a good example of functional tasks which can be performed by exploiting the I\&F response of an RTD-LD node. 

\subsection{Feedforward network of optoelectronic nodes}
\begin{figure*}[ht]
\centering
\includegraphics[width=0.99\textwidth]{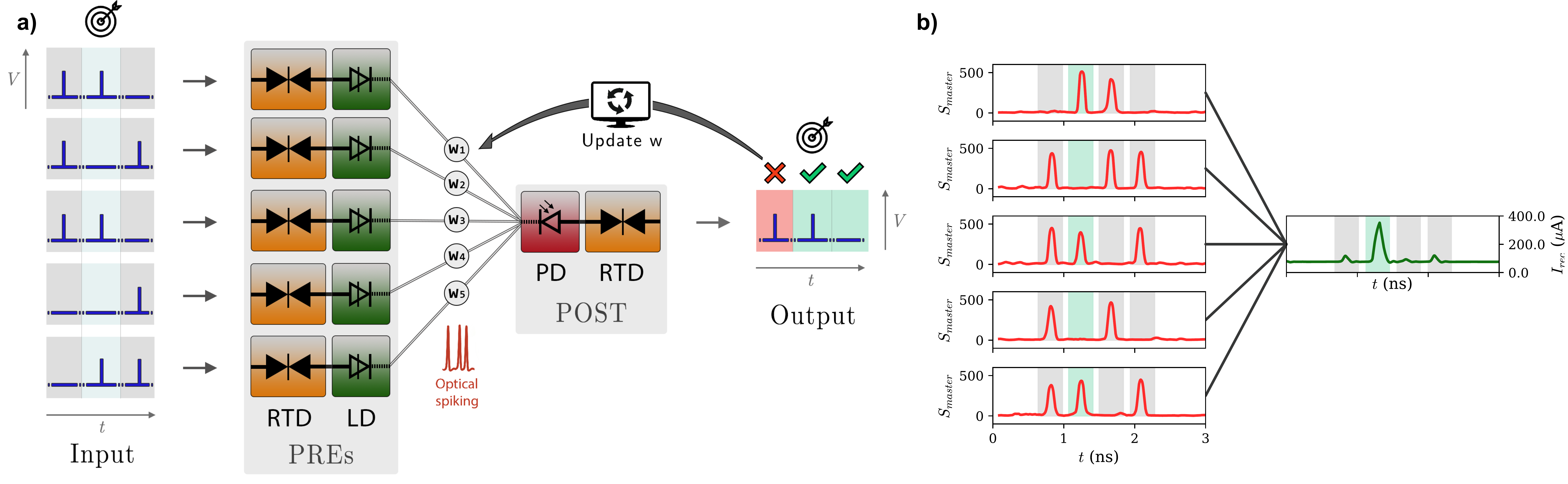}
	\caption{\label{fig:networks}\emph{a)} Network architecture diagram, illustrating how patterns of input electronic pulses (in blue) enter the RTD-LD nodes and are propagated as optical signals to the downstream node using weighted connection. The output state of the downstream node is compared to the label, and if there is a mismatch between label and output state, the weights are updated. Desired pattern is highlighted with the target icon. \emph{b)} Visualization of inference in 5-to-1 feedforward network numerical model. The guiding signals representing pattern labels are visualized as background shading (green for 'True', grey for 'False'). Only a particular spatial pattern ([1 0 1 0 1], green) results in firing of electric spike of the downstream RTD-PD node (green current trace). The red timetrace represents a simple moving average of the LD output optical signal over $t_{MA} =$\SI{0.1}{\nano\second}.}
\end{figure*}
Since the information processing capability of an artificial neural network (ANN) typically grows with increasing network complexity, demonstrating networking performance with multiple optoelectronic spiking nodes is of key importance. Here, we numerically investigate the operation of a spiking variation on the single layer, feedforward perceptron model with all-to-one layout. Such network processes input spike-represented data by weighting the signal from each upstream node and summing up all the weighted inputs on the the downstream node, which fires a spike if the weighted input sum exceeds the firing threshold. In the demonstrated model, the spatiotemporal patterns of input superthreshold stimuli are injected into the first layer of neurons (PREs), where each stimuli results in a guaranteed optical spiking outcome from the corresponding PRE node. The optical spiking signals from the PREs are weighted by attenuating them (multiplying their intensity by a given factor $w_n$ in the numerical model). During the network learning phase, a guiding signal carries the data labels alongside each pattern, marking it as wanted (True) or unwanted (False) via change in amplitude. The downstream POST neuron performs the temporal integration of the upstream inputs and fires a spike if the voltage of spiking threshold is surpassed (integrate \& fire operation). A diagram of the network is depicted in Fig. \ref{fig:networks}a, showing how different patterns (consisting of spikes, in blue) may result in activation of spikes and illustrating the dependence of weights on the output of the downstream node.

In particular, the investigated network consist of five layer 1  RTD-LD nodes (PREs, biased in the valley (in PDR II), $V_{DC}=$\SI{770}{\milli\volt}), whose output optical signals are propagated through unidirectional, feed-forward links (each with weight $w_i$) to a single, layer 2 PD-RTD (POST) node biased in the valley (in PDR II). In the PD-RTD, the PD is current-coupled into the spiking RTD element (with the PD conversion factor $\kappa$), directly converting the incoming optical intensity into the electrical domain and resulting in activation of an electronic spiking signal. In the PREs, we utilize superthreshold input trigger pulses of length $t_{pulse}=$ \SI{80}{\pico\second}, resulting in excitatory (increasing intensity) optical pulses. Since the output current of a PD at a given time directly depends on the input light intensity and temporal distance from previous optical spikes, only certain weighted pulse patterns may result in sufficiently strong current modulation, activating a spike in the downstream node. That is the working principle of the network model for input spatiotemporal spike-pattern recognition. Visualization of the pattern recognition in the network is shown in Fig. \ref{fig:networks}b. In this network, the temporal separation between each 5-bit input pattern is set to \SI{420}{\pico\second}, corresponding to full network processing capacity of \SI{11.9}{\Gbps}.

\subsection{Networks: supervised learning method for spatiotemporal pattern recognition}

Training algorithms are fundamental for useful utilization of artificial neural networks (ANNs). However, training methods for spiking neural networks (SNNs) differ from those used for conventional ANNs, which are typically based on backpropagation \cite{Pfeiffer2018}. SNNs can be trained using either biologically-plausible local learning rules (e.g. spike-timing dependent plasticity (STDP), long-term potentiation) or using other specially designed algorithms such as ReSuMe \cite{Ponulak2010}, Resilient-Back-Propagation (RProp) inspired supervised learning \cite{McKennoch2006} and SuperSpike \citep{Zenke2018}, among others. In this work, we introduce an offline supervised learning rule, following the approach introduced in \cite{Wang2018} for training memristor-based neural networks. However, in contrast to \cite{Wang2018}, our system propagates information using optical spike trains, allowing us to fully benefit from the advantages of optical signalling (e.g. high-bandwidth, low loss waveguiding, non-interacting signals etc). Data processing in our network follows the two typical phases: a) \textit{training} phase and b) \textit{inference} phase. During the training phase, labelled patterns are processed by the network. By comparing the output state of the network with the label, according adjustments are made to the network weight matrix. The learning phase consists of multiple epochs, and progresses until the weights stabilize. During a single epoch, the dynamical evolution of all the RTD-based nodes in the network is numerically evaluated. The use of teacher signals (which carry the label of the pattern) allows for processing of multiple patterns in a single epoch. In the learning phase, three independent patterns are processed per single epoch of $t=$ \SI{5}{\nano\second}.

\begin{figure}[ht!]
\centering
\includegraphics[width=0.42\textwidth]{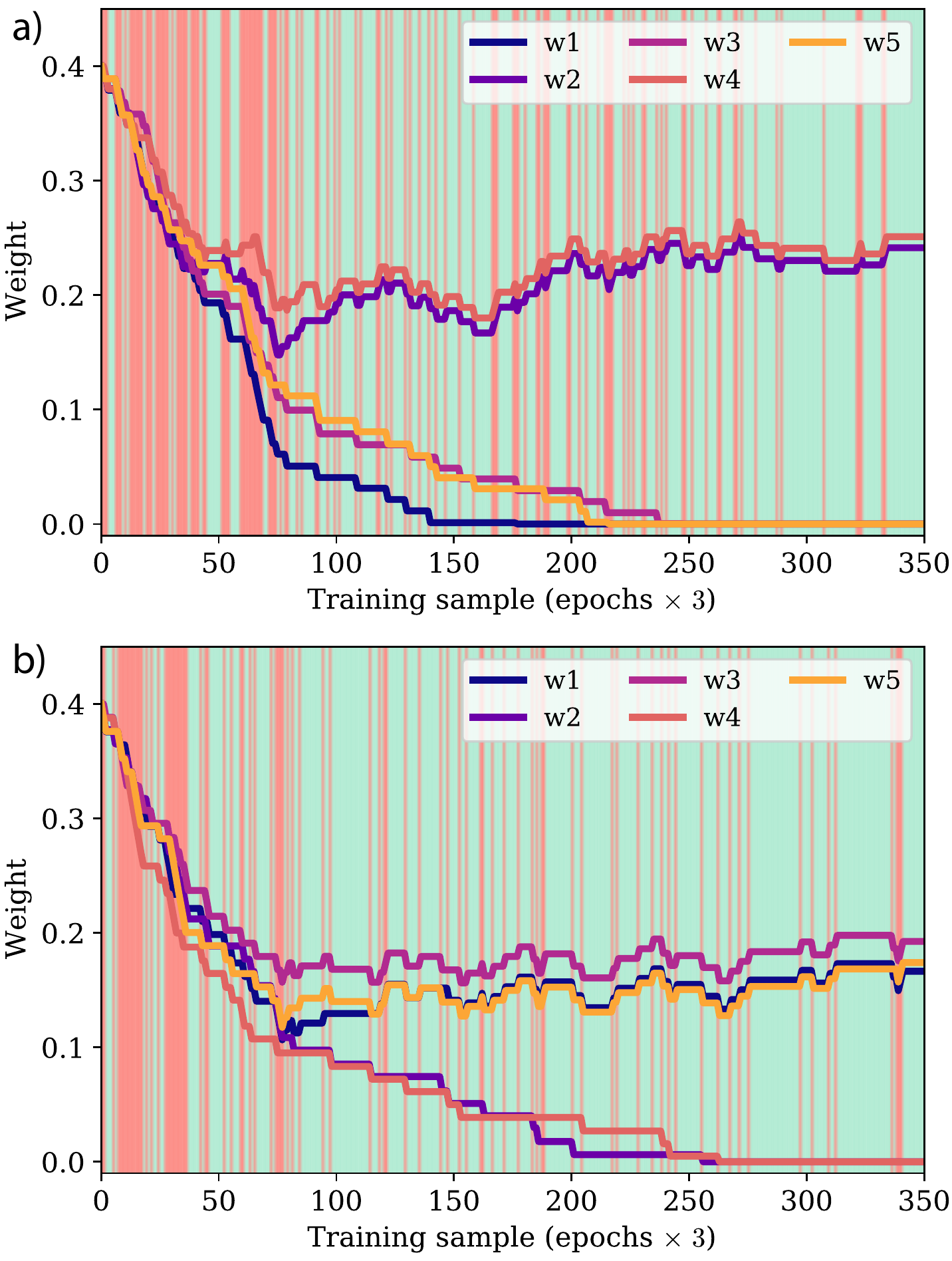}
	\caption{Demonstration of the supervised learning process for two different spatial patterns with varying number of active bits: \emph{a)} [0 1 0 1 0] and \emph{b)} [1 0 1 0 1]. As the system is used to process to labelled patterns in each epoch, the weights are adjusted using the local learning rule, strengthening connections which produced false negative results and weakening links which produced false positive results. The background colour shows network state (True/False) during each step.}\label{fig:learndiagram}
\end{figure}

\begin{figure}[ht]
\centering
\includegraphics[width=0.39\textwidth]{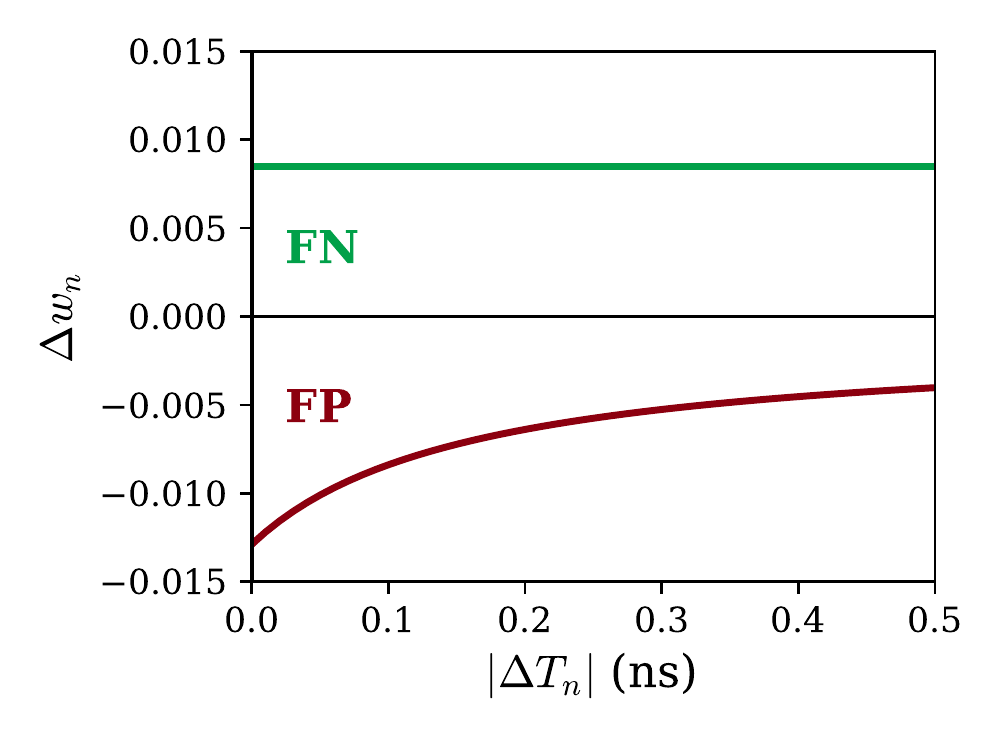}
	\caption{Weight adjustment factor as a function of POST-PRE spike temporal distance $|\Delta T_n|$. For false negative (FN, in green) cases, the weight adjustment is a constant fixed value. For false positive, the magnitude of weight adjustment is a function of $|\Delta T_n|$, with close spikes being depotentiated more strongly.}\label{fig:STDP_curve}
\end{figure}

Fig. \ref{fig:learndiagram} shows the learning process. The target input is a 5-bit spatial pattern, either [0 1 0 1 0] (Fig. \ref{fig:learndiagram}a) or [1 0 1 0 1] (Fig. \ref{fig:learndiagram}b), and the network is initiated with all weights set to an initial value $w=0.4$. We want to note here that the weights depend on the current conversion factor $\kappa$ of the PD, which was selected in this demonstration to bound the weights in the usual interval [0,1]. During each learning epoch, three random patterns are picked, with a probability $P_{t}=0.25$ of picking the target and $P_{f}=0.75$ of picking any other pattern. Fig. \ref{fig:learndiagram} shows the evolution of the weights during each learning step. Green background represents \textit{True} outcomes (\textit{True positive}, \textit{True negative}) while red represents \textit{False} outcomes (\textit{False positive}, \textit{False negative}). For either \textit{True} output state, no weights are adjusted during the learning step. For the \textit{False positive} output state, the weights that contributed to the firing are weakened, with $\Delta w$ being a function of PRE-POST spike separation. The closer the PRE node's spike was to activation of a \textit{False positive} POST spike, the higher is the depotentiation (weakening) effect. This is a supervised variation on the STDP learning protocol, a specific kind of Hebbian learning approach which is believed to constitute part of the learning process in biological neural networks. A simple rational function was selected for the weight adjustment:

\begin{equation}
    \Delta w_n= \frac{a}{b\cdot|\Delta T_n|+c}+d
\end{equation}
where
\begin{equation}
    \Delta T_n = T_{POST}-T_{PRE,n}
\end{equation}
represents the time interval between the spikes from the POST and the PRE neuron $n$, $a=9.35\cdot10^{-3}$, $b=5\cdot10^{9}$, $c=0.8$, $d=1.5\cdot10^{-3}$. The numerical coefficients in the rational function were selected based on observed distances between spikes in PRE-POST neurons and the corresponding desired weight adjustments. Weight adjustment factors as a function of spike separation time can be seen in Fig. \ref{fig:STDP_curve}.

\begin{figure}[ht]
\centering
\includegraphics[width=0.40\textwidth]{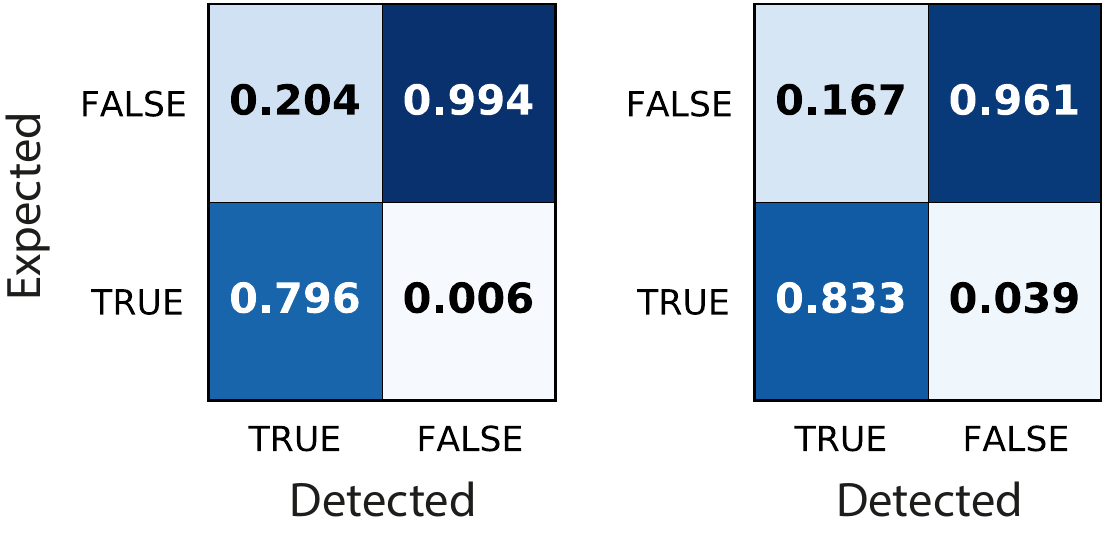}
	\caption{\emph{(Left)} Confusion matrix for inference of [0 1 0 1 0] against all other pattern with two ON bits ($n=10$ different patterns, $540$ total inference steps). \emph{(Right)} Confusion matrix for inference of pattern [1 0 1 0 1] against all other pattern with three ON bits ($n=10$ different patterns, $540$ total inference steps). }\label{fig:learnaccuracy}
\end{figure}

As the training process proceeds, the occurrence of false outcomes gets more and more rare. For both tested patterns, the system reaches a stable weight setting in approximately 300 patterns (100 epochs). This network implementation utilizes only positive weight values, making the solution physically feasible. After the training phase, the network can perform inference for recognition of the selected spatiotemporal 5-bit pattern. We tested all patterns with equal number of active bits against a single desired target pattern: [0 1 0 1 0] in one measurement, and [1 0 1 0 1] in the other. When testing inference accuracy for [0 1 0 1 0] against all patterns with $n_{ON}=2$ active bits, the total True response accuracy (with $540$ inferred patterns) was $97.4\%$. Inferring the pattern [1 0 1 0 1] against all patterns with same number of ON bits ($n_{ON}=3$) in $540$ inference steps yields total True response accuracy of $94.8\%$. The confusion matrices for both of these inference procedures are shown in Fig. \ref{fig:learnaccuracy}.

\newpage
\section{Conclusions}
In this work, we introduce a nano-optoelectronic neuromorphic node based on the DBQW-based resonant tunnelling diode exhibiting regions of negative differential conductance (NDC), enabling neuron-like electronic spiking responses at over GHz rates. The nodes consist of this highly nonlinear, high bandwidth RTD element coupled to either a high-sensitivity photodetector or a nanoscale laser to enable the reception and transmission of optical spikes, respectively. These devices offer multiple desirable properties such as suitability for either hybrid or monolithic integration, highly reduced footprint, high-speed operation (\SI{<100}{\pico\second} long input signals) and low energy requirements (operation with mV trigger pulse amplitudes and energies of $<$pJ/spike). All these aspects are crucial for the future development of on-chip integrated neuromorphic photonic computing platforms. We investigate and analyze the dynamical behaviour of the proposed neuromorphic optelectronic system and discuss feasible hardware implementations of individual nodes as well as nodes in interconnected network architectures. We also numerically demonstrate functional information processing tasks, including 8-bit pattern recognition and image edge-feature detection at over \SI{10}{\Gbps} rates (using \SI{50}{\pico\second} ps long input signals). Finally, we demonstrate network operation, investigating a 5-to-1 feedforward (two-layer) spiking neural network architecture. Using physical models for each node, we demonstrate that the numerically implemented network can be used to classify spatial 5-bit pulse patterns encoded in time, and we propose a supervised learning scheme that employs a spike-timing dependent learning rule. During the inference phase, we demonstrate $94\%+$ accuracy for spatial pulse pattern recognition. To the best of our knowledge, the reported results are the first theoretical demonstration of RTD-based neuromorphic optoelectronic spike information processing, delivering successful operation in key tasks (e.g. pattern recognition, image edge-detection) by utilizing either implementation-friendly single node solution, or using multiple interconnected devices in the form of a photonic feed-forward spiking neural network. 

\section{Acknowledgements}
The authors acknowledge funding support from the European Commission (Grant No. 828841 ChipAI-H2020-FETOPEN-2018–2020), the UKRI Turing AI Acceleration Fellowships Programme (Grant No. EP/V025198/1), and the Office of Naval Research Global (Grant No. ONRGNICOPN62909-18-1-2027).

\newpage
\bibliography{library,libUIB}

\newpage
\onecolumngrid
\begin{center}
\section*{Supplementary Information}
\end{center}
\twocolumngrid

\setcounter{equation}{0}
\setcounter{figure}{0}
\setcounter{table}{0}
\setcounter{section}{0}
\setcounter{page}{1}
\makeatletter
\renewcommand{\theequation}{S\arabic{equation}}
\renewcommand{\thefigure}{S\arabic{figure}}
\renewcommand{\thepage}{Supp-\arabic{page}}

\section{RTD I-V characteristics and simulation parameters}
The RTD I-V characteristics used in this work has an analytical form of \citep{SDC-EDL-96}:
\begin{align}
	f\left(V\right) & =a\ln\left(\frac{1+e^{\frac{q_{e}}{k_{B}T}\left(b-c+n_{1}V\right)}}{1+e^{\frac{q}{k_{B}T}\left(b-c-n_{1}V\right)}}\right)\label{eq:I-V_curve}\\
	& \times\left(\frac{\pi}{2}+\tan^{-1}\left(\frac{c-n_{1}V}{d}\right)\right)+h\left(e^{\frac{q_{e}}{k_{B}T}n_{2}V}-1\right).\nonumber 
\end{align}
Here, $T$ is the temperature, $q_{e}$ is the electron charge and $k_{B}$ is the Boltzmann's constant. The parameters $a$, $b$, $c$, $d$, $n_{1}$, $n_{2}$, $h$ depend on the geometry of the barrier and the confinement energy levels. The following parameter values were found after fitting experimental data: 
\begin{table}[h!]
	\begin{tabular}{@{}ll@{}}
		\toprule
		Parameter &  Value               \\
		\midrule
		$a$       & $-5.5\times10^{-5}\,\text{A}$ \\
		$b$       & $0.033\,\text{V}$             \\
		$c$       & $0.113\,\text{V}$             \\
		$d$       & $-2.8\times10^{-6}\,\text{V}$ \\
		$n_{1}$   & $0.185$                       \\
		$n_{2}$   & $0.045$                     \\
		$h$       & $18\times10^{-5}\,\text{A}$  
	\end{tabular}
\end{table}

The use of negative parameter values is justified as Eq. \ref{eq:I-V_curve} is derived from an idealised system that has been fitted to a realistic one adjusting the parameter values. The electronic parameters of the RTD circuit are:
\begin{table}[h!]
	\begin{tabular}{@{}ll@{}}
		\toprule
		Parameter     & Value                            \\ \midrule
		$R$           & $10\,\Omega$                               \\
		$L$           & 126$\times$10$^{-9}$\,H     \\
		$C$           & 2$\times$10$^{-15}$\,F      \\
	\end{tabular}
\end{table}

\subsection{Self-oscillations and slow-fast dynamics\label{subsec:self-oscillations}}
It is well-known that an RTD driven by an electrical DC source exhibits self-oscillations when biased in the NDC region, and a steady response (given by the intersection between the load line and the RTD I-V curve) when biased in either PDC region \citep{WAW-IEEE-18,ISA-IEEE-17,DNN-IEEE-17}.
From the analytical point of view, this notion is valid (although not exact) provided that the load line is "almost vertical" and
the coefficient $\mu=\text{\ensuremath{\sqrt{C/L}}}$ is much smaller than the absolute value of the minimal differential conductance in the NDC region \citep{RFJ-CHAOS-18,RJI-OE-13,OPRJ-PRApp-20}. The first condition can be assumed in our study, since the slope of the load line (i.e., $-1/R=-0.1\,\Omega^{-1}$) is much bigger in magnitude than that of the segment that connects the peak and the valley ($\Delta I/\Delta V=-2.39\times10^{-3} \Omega^{-1}$). The second condition applies as well as the minimal value of $f'(V)$ in the NDC region is $-5.5\,\Omega^{-1}$while $\mu=1.26\times10^{-4}\,\Omega^{-1}$.
\begin{figure}[t]
	\centering{}\includegraphics[width=1\columnwidth]{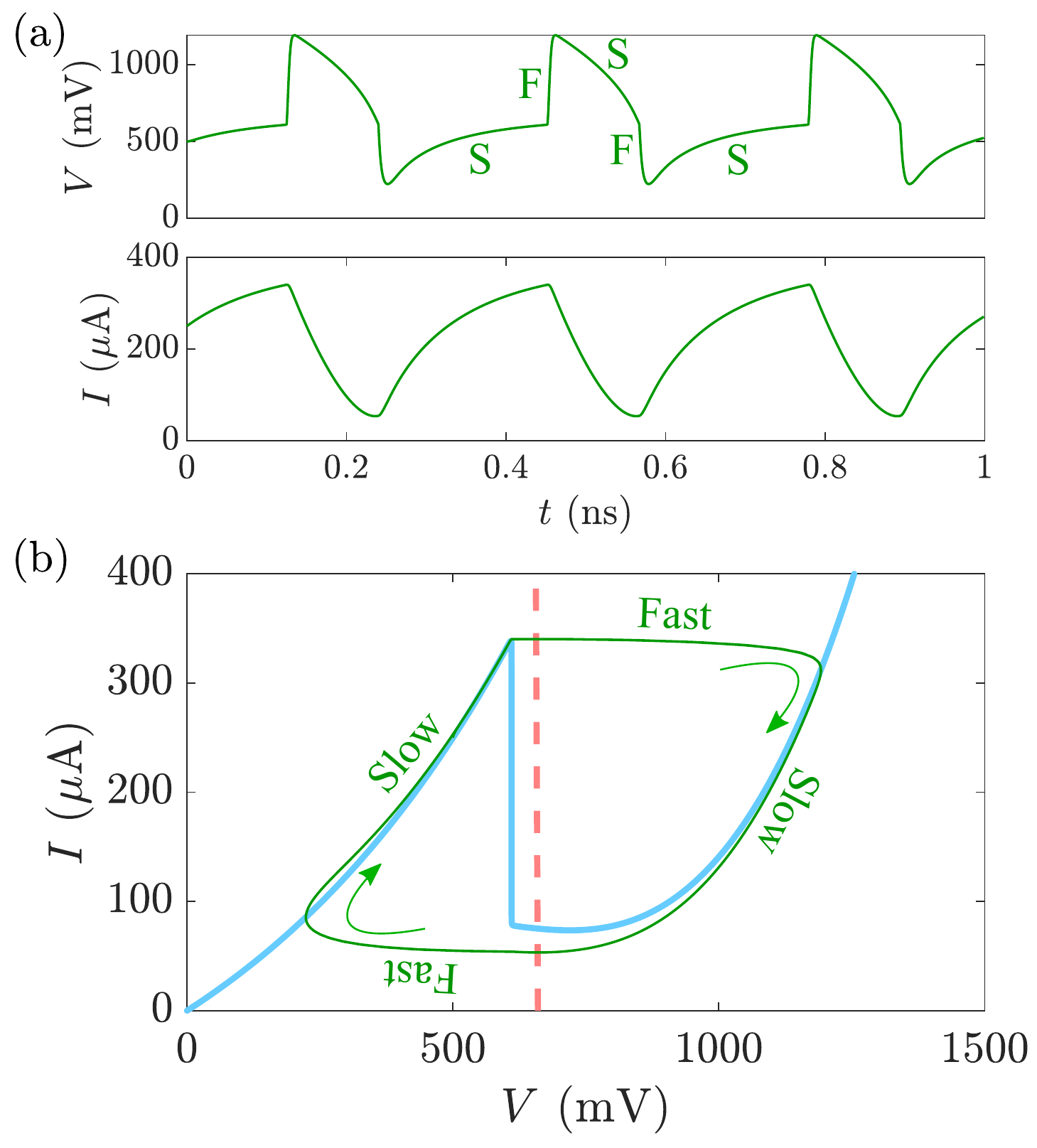}
	\caption{\label{fig:slowfast}\emph{a}) Self oscillations time series of the RTD biased at $V_{0}$ = \SI{660}{\milli\volt}. \emph{b}) Orbit of the self oscillations on the phase plane (straight green line) together with the I-V characteristic (straight blue line) and the load line (dashed red line). Stages of slow and fast dynamics are also shown.}
\end{figure}
The coefficient $\mu$ also accounts for the stiffness in the dynamics. Provided that $\mu$ is much smaller in magnitude than the minimal conductance, the self-oscillations have sharp-cornered orbits and each period has two stages of slow and two stages of fast dynamics \citep{Ortega-Piwonka2021}, as illustrated in Fig. \ref{fig:slowfast}. Each slow stage takes place at each PDC region, where the orbit remains close to the I-V curve until reaching either the peak or the valley. Here, the orbit makes a fast jump into the other PDC region, the voltage $V$ changes abruptly in a very short time, with a slight change in the current $I$, which does not exhibit the fast stages. Consequently, the RTD generates current and voltage spikes periodically with a frequency that depends on the bias voltage \citep{OPRJ-PRApp-20} but remains at about \SI{3}{\giga\hertz}.
\subsection{Excitability\label{subsec:Excitability}}
The concept of \emph{excitability} was originally coined to refer to the capacity of living organisms to respond to an external stimulus, provided that such stimulus is above a certain threshold \citep{HH-JOP-52,HH2-JOP-52}, but it is also applicable in other areas such as image processing \citep{kuhnert89}, neuron-like semiconductors \citep{SNJ-JAP-11} and signal generation \citep{GHR-PRL-07,SBB-PRL-14,BKY-OL-11}. A physical system is referred to as excitable if, after a sufficiently strong (super-threshold) perturbation, it responds with a long, complex trajectory before returning to its natural state of equilibrium. While this response takes place, the system is unable to respond to any other perturbation, strong or weak. The duration of the response is thus known as refractory (lethargic) time. For a perturbation below the threshold, the response is a fast, exponential decay, typical of linear systems. Excitable systems are particularly interesting in the context of spiking because the response to a super-threshold perturbations can be interpreted as a spike or pulse, while unresponsiveness to weak perturbations confers them a robust character.
\begin{figure}[t!]
	\centering{}\includegraphics[width=1\columnwidth]{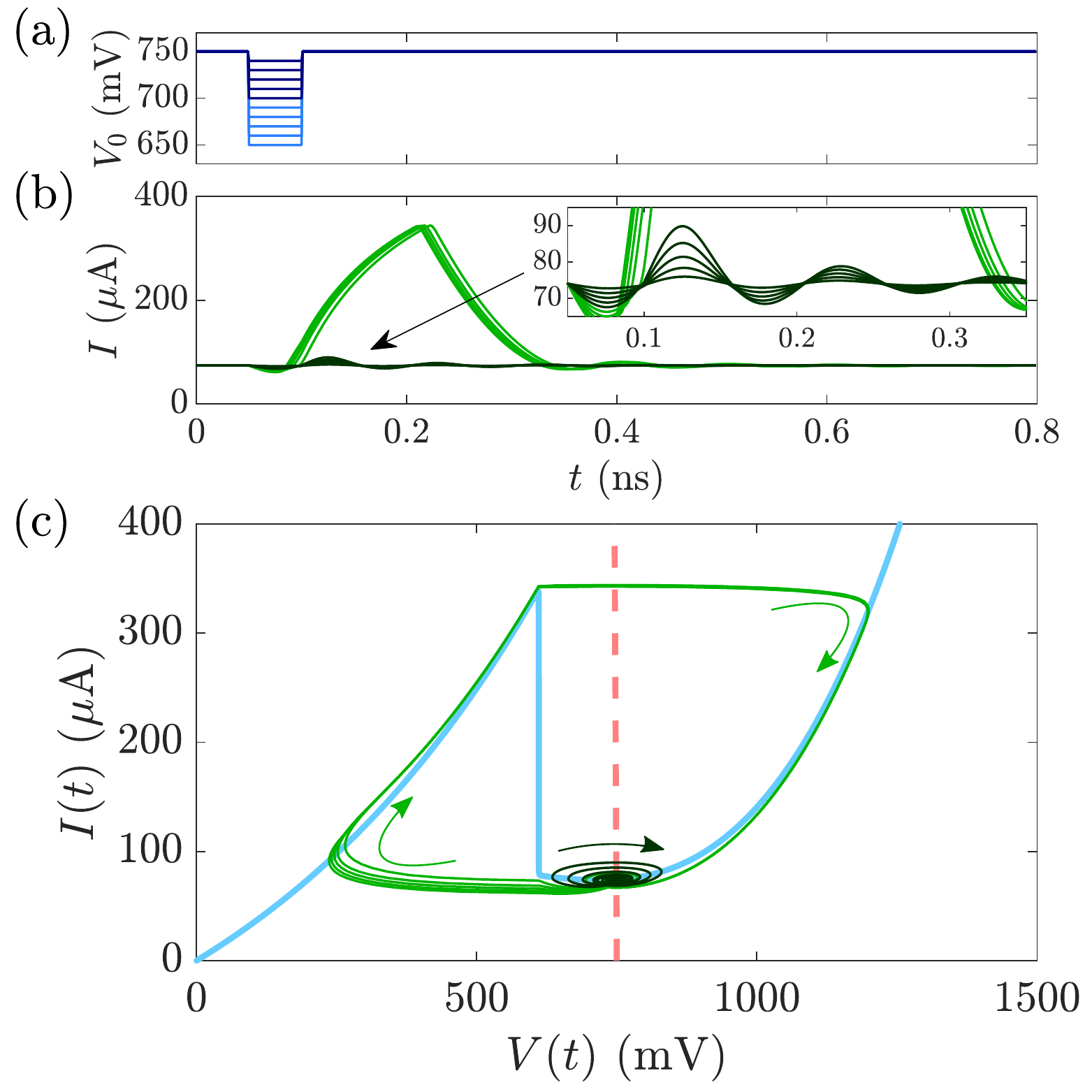}
	\caption{\emph{a}) Downwards square voltage pulses used to perturb the RTD circuit biased at $V_{0}$ = \SI{750}{\milli\volt}. \emph{b}) Current-responses of the circuit over time. The inset zooms over the responses to below-threshold perturbations. \emph{c}) Responses of the circuit on the phase plane.}\label{fig:excitable}
\end{figure}
When the circuit is biased at $V_{0}$ = \SI{750}{\milli\volt} (i.e., in the second PDC region but close to the NDC region), it remains steady at about $I$ = \SI{74}{\micro\ampere}. Fig. \ref{fig:excitable} shows the response of the circuit to inhibitory (i.e., downward) square voltage pulses of \SI{50}{\pico\second} long and depths ranging from 10 to \SI{100}{\milli\volt}. If the pulse depth is \SI{50}{\milli\volt} or less, the circuit responds with low-amplitude relaxation oscillations and decays back into the steady state in a relatively short time. However, if the pulse depth is larger than \SI{50}{\milli\volt} (excitability threshold), the circuit responds with a large current pulse (about \SI{280}{\micro\ampere} in amplitude and \SI{250}{\pico\second} long) before returning to the steady state. The voltage across the RTD is also affected and the circuit's response follows a phase-space orbit reminiscent of that of the self-oscillations that arise when the circuit is biased in the NDC. The slow and fast stages of the periodic self-oscillations are also present here.
\section{Laser diode simulation parameters}
Given the dimensions of the device under study, the nanoscale LD model utilized in this work uses the following set of parameters:

\begin{table}[h!]
	\begin{tabular}{@{}ll@{}}
		\toprule
		Parameter     & Value                            \\ \midrule
		$N_{0}$       & $5\times10^{5}$                            \\
		$\gamma_{m}$  & $10^{7}\,\text{s}^{-1}$                    \\
		$\gamma_{l}$  & $10^{9}\,\text{s}^{-1}$                    \\
		$\gamma_{nr}$ & $2\times10^{9}\,\text{s}^{-1}$             \\
		$\tau_{p}$    & 5$\times$10$^{-13}$\,s      \\
		$\kappa$      & 1.1$\times$10$^{-7}$\,A     \\
		$\eta$        & 1                                          \\
		$J$         & 210\,$\mu$A                 \\
	\end{tabular}
\end{table}

Thus, the Purcell enhanced factor is $\beta=\gamma_{m}/(\gamma_{m}+\gamma_{l})=0.099$, proper of nanoscale lasers. When injected with a constant current, laser diodes emit with an intensity that increases in proportion to such current, provided that the current is above the lasing threshold. Indeed, the stable equilibrium photon number is a piecewise linear function of the input bias current,

\begin{equation}
	S=\begin{cases}
		0 & J\leq J_{tr},\\
		\tau_{p}\left(\frac{J}{q_{e}}-\left(\gamma_{l}+\gamma_{m}+\gamma_{nr}\right)\left(N_{0}+\frac{1}{\gamma_{m}\tau_{p}}\right)\right) & J\geq J_{tr},
	\end{cases}\label{eq:node_ss_S}
\end{equation}

where $$J_{tr}=q_{e}\left(\gamma_{l}+\gamma_{m}+\gamma_{nr}\right)\left(N_{0}+1/\gamma_{m}\tau_{p}\right)=337.54\,\text{\ensuremath{\mu\text{A}}}$$ for the parameters chosen in this study.

\end{document}